\documentstyle[11pt,conf_iap]{article}
\begin{document}

\heading{THE FUTURE OF MASSIVE VARIABILITY SEARCHES
\footnote{Invited talk, presented at the 12th IAP Colloquium: Variable Stars
and the Astrophysical Returns of Microlensing Surveys.}
}

\author{BOHDAN PACZY\'NSKI}
       {Princeton University Observatory, Princeton, NJ 08544--1001, USA; \\
        E-mail: bp@astro.princeton.edu}

\bigskip

\begin{abstract}{\baselineskip 0.4cm}
This is a personal review of various issues related
to massive photometric and astrometric searches.
A complete inventory of variable stars down to almost any magnitude
limit will improve our understanding of the stellar
evolution and the galactic structure.  A search for detached
eclipsing binaries will improve the distance scale, the
value of the Hubble constant, and the
age of the oldest stars.  A search for supernovae will help
the determination of cosmological parameters $ \Omega $ and $ \Lambda $.
A search for microlensing events will provide insight into the stellar
mass function, dark matter, and may lead to a discovery of earth-mass
planets.
\end{abstract}

\section{Introduction}

Many of the most interesting new results in astrophysics
were obtained with survey projects, opening up a new range
of photon energies, lower flux limits, more precise position
measurements, etc.  A rapid increase of the power
of computers, and a rapid decrease of their cost, opened
up yet another dimension for the surveys: the number of objects
that can be measured and monitored.  The massive searches
for gravitational microlensing have lead to the detection
of over 100 microlensing events by the MACHO, OGLE, DUO, and EROS
collaborations (Paczy\'nski 1996b and references therein).  These
projects demonstrated that automated real time processing of photometric
measurements of up to $ \sim 10^7 $ stars every night is possible with
very modest hardware and operating costs.  They also demonstrated
that while the original goal of finding the very rare microlensing
events has been reached, much more has been accomplished;
a very large diversity of scientific results was obtained,
related to the galactic structure and stellar variability.
New programs have been stimulated,
providing new ways to improve the cosmic distance scale,
the age estimates of the oldest stellar systems, a robust
determination of the stellar luminosity function and mass function.

The microlensing searches
concentrate on small selected areas, the galactic bulge and the
Magellanic Clouds, and they reach down to $ \sim 20-21 ~ mag $ stars.
Note that the total number of all stars over the whole sky which
are brighter than $ 15 ~ mag $ is $ \sim 2 \times 10^7 $ (Allen 1973),
of which only $ \sim 50\% $ are visible from any given site on any given
night.  Therefore, the processing power comparable to that of
the existing microlensing searches can allow nightly photometric
measurements of all stars brighter than $ \sim 15 ~ mag $.
An incremental increase in the data acquisition and data processing rates
can gradually bring into the monitoring programs all
objects brighter than magnitude 16, 17, etc.  At every step there
are new and interesting scientific issues to address.  

Some recent and planned large scale optical surveys and
searches are described in the next section.
Some examples of possible scientific programs feasible at various
magnitude limits are described in section 3, and some possible technical
implementations are described in section 4.

\section{The Past and Current Massive Monitoring Programs}

Most variable stars which can be found in the catalogues were discovered
with massive photographic monitoring programs, carried out over many
decades at Harvard College Observartory, Sonnenberg Observatory, and
many other observatories.  The majority of supernovae
were discovered with photographic plates or films, even though many
were also discovered visually or with CCD detectors.  

Recently, a number of massive photometric searches for gravitational
microlensing events were began with either photographic or CCD
detectors: DUO (Alard 1996b), EROS (Aubourg et al. 1993),
MACHO (Alcock et al 1993), and OGLE (Udalski et al. 1992).  These
projects not only revealed a number of microlensing events (Alcock
et al. 1993, 1995a,c, Pratt et al. 1996, Aubourg et al. 1993, Alard 1996b,
Udalski et al. 1993, 1994a), but also a huge number of variable stars
(Udalski et al. 1994b, 1995a,b, Alard 1996a, Alcock et al. 1995b, Grison
et al. 1995, Cook et al. 1996, Ka\l u\.zny et al. 1995a).
The total number of variable stars in 
the data bases of these four collaborations can be estimated to be about 
$ 10^5 $, most of them new.  

A very important aspect of these projects is
their very large scale, so the search for variable
objects has to be done, and is done, with computer software, following
some well defined algorithms.  This means that the sensitivity of the
searches to detecting variables of any specific kind can be well
calibrated.  So far a preliminary calibration was done only for
the sensitivity to detect microlensing events (Udalski et al. 1994a,
Alcock et al. 1995b).  A similar procedure can be done, and certainly
will be done, to determine the sensitivity to detect variable stars
of various types, magnitudes, amplitudes, periods, etc.  This is a new
possibility, rarely if ever available to the past photographic searches.

In addition to massive searches there are a number of on-going observing
programs of selected groups of known variables, in order to understand
their long term behavior.  Many such observations are done with
robotic telescopes, looking for supernovae, optical flashes related
to gamma-ray bursts, near earth asteroids, and many other variable
objects (Hayes \& Genet l989, Honeycutt \& Turner 1990, 
Baliunas \& Richard l991, Filippenko 1992, Perlmutter et al. 1992,
Schaefer et al. 1994, Akerlof et al. 1994,
Hudec \& Soldan 1994, Kimm et al. 1994, Henry \& Eaton 1995).  
Many robotic telescopes are operated by amateur astronomers.
The CCD cameras and computers became so inexpensive that they can be
afforded by non-professional astronomers, or by groups of non-professional
astronomers.

One of the most important massive observational programs is being done with 
the satellite Hipparcos, which will provide excellent astrometry and 
photometry for the brightest $ 10^5 $ stars over the whole sky, and less 
accurate data for another one million stars, as described in many articles 
published in the whole issue of Astronomy and Astrophysics, 
volume 258, pages 1 -- 222.

There are many new massive searches being planned by many different groups.
These include people who are interested in detecting possible optical 
flashes from cosmic gamma-ray bursts (Boer 1994, Otani et al. 1996), 
people who search for supernovae, general variable stars, near-earth
asteroids, distant asteroids, Kuiper belt comets, etc.  Perhaps the
largest scale survey project is the Sloan Digital Sky Survey (Gunn \&
Knapp 1993, Kent 1994).  The All Sky Patrol Astrophysics (ASPA, Braeuer \& 
Vogt, 1995) is the project to replace the old photographic sky monitoring
with the modern CCD technology.  It is not possible
to list all the proposed programs, as there are so many of them.  They have
a large variety of scientific goals, but they all have one thing in
common: they are all aimed at the automatic photometry and/or astrometry
of a huge number of objects on a sustained basis, and most of them are
proposing a search for some rare, or extremely rare type of objects or
events.  

\section{Scientific Goals}

There are different types of scientific results which will
come out of any major survey, including the future all
sky monitoring programs.  

{\bf First}, if a survey is done
in a systematic way which can be calibrated, then it will
generate large complete samples of many types of objects:
ordinary stars of different types, eclipsing binaries,
pulsating stars, exploding stars, stars with large proper
motions, quasars, asteroids, comets, and other types of
objects.  Such complete samples are essential for
statistical studies of the galactic structure, the stellar evolution,
the history of our planetary system, etc.  

{\bf Second}, the identification
of more examples of various types of objects will make it possible
to study them in great detail with the follow-up dedicated
instruments and will help to improve the empirical calibration of
various relations.  Bright detached eclipsing binaries and
bright supernovae are just two examples of objects which call
for the best possible calibration.  

{\bf Third}, some very rare objects
or events will be detected.  Some may uniquely assist us in 
understanding critical stages of stellar evolution.  A spectacular
example from the past is FG Sagittae, a nucleus of a planetary
nebula undergoing a helium shell flash in front of our telescopes
(Woodward et al. 1993, and references therein).
Some may provide spectacles which bring astronomy to many
people.  A few recent examples are the supernova 1987A in the Large
Magellanic Cloud, a collision of the comet Shoemaker-Levy
with Jupiter in the summer of 1994, and the bright comet
Hyakutake in the spring of 1996.

{\bf Fourth}, fully automated real time data processing will provide
instant alert about variety of unique targets of opportunity:
supernovae, gravitational microlensing events, small asteroids
that collide with earth every year, etc.  Such alerts will provide
indispensable information for the largest and most expensive space
and ground based telescopes, which have tremendous light collecting 
power and/or resolution but have very small fields of view.

{\bf Fifth}, the archive of photometric measurements will provide
a documentation of the history for millions of objects, some of which
may turn out to be very interesting some time in the future.  The
Harvard patrol plates and the Palomar Sky Survey atlas provide an
excellent example of how valuable an astronomical archive can be.

{\bf Sixth}, some unexpected new objects and phenomena may be discovered.
There is no way to know for sure, but it is almost always the case
that when the amount of information increases by an order of magnitude
something new is discovered.

One may envision beginning the all sky variability survey with very low
cost equipment: a telephoto lens attached to a CCD camera, with the cost
at the level of a personal computer.  A ''low end'' system can easily 
record stars as faint as 14 magnitude.  Naturally, many
small units are needed to monitor the whole sky.  

It is useful to realize how many stars there are in the sky as a function
of stellar magnitude.  This relation is shown with a solid line in Figure 1,
following Allen (1973).  There are $ 10^3 $, $ 10^4 $,
$ 10^5 $, $ 10^6 $, $ 10^7 $, $ 10^8 $ stars in the sky
brighter than approximately $ 4.8, ~ 7.1, ~ 9.2, ~ 11.8, ~ 14.3, $ and
$ 17.3 ~ mag $, respectively.
Also shown in the same Figure are the numbers of 
know binaries of three types: Algols, contact binaries (W UMa), and 
binaries with spotted companions (RS CVn).  Note that among the brightest
stars the fraction of these binaries is very high, presumably because
these stars, with $ m < 6 ~ mag $, are studied so thoroughly.  When
we go faint the incompleteness sets in.  

In the following sub-sections I shall discuss a few specific types of
variables, pointing out to various indicators of the incompleteness.
All numbers are taken from the electronic edition of the 4th General
Catalogue of Variable Stars as available on a CD ROM (Kholopov et al. 1988).

\subsection{RS Canum Venaticorum Binaries}

The incompleteness is most dramatically apparent for the
RS CVn type binaries, where very few stars fainter than 5th magnitude
are known.  These variables have small amplitudes, typically
only 0.1 or 0.2 $ mag $, so they cannot be found with photographic
searches.  A systematic CCD search should reveal lots of such variables
in the magnitude range $ 6 - 10 $, and of course even more among
fainter stars.  Periods are typically a few days to a few weeks.


\begin{figure}[p]
\vspace{8cm}
\includegraphics{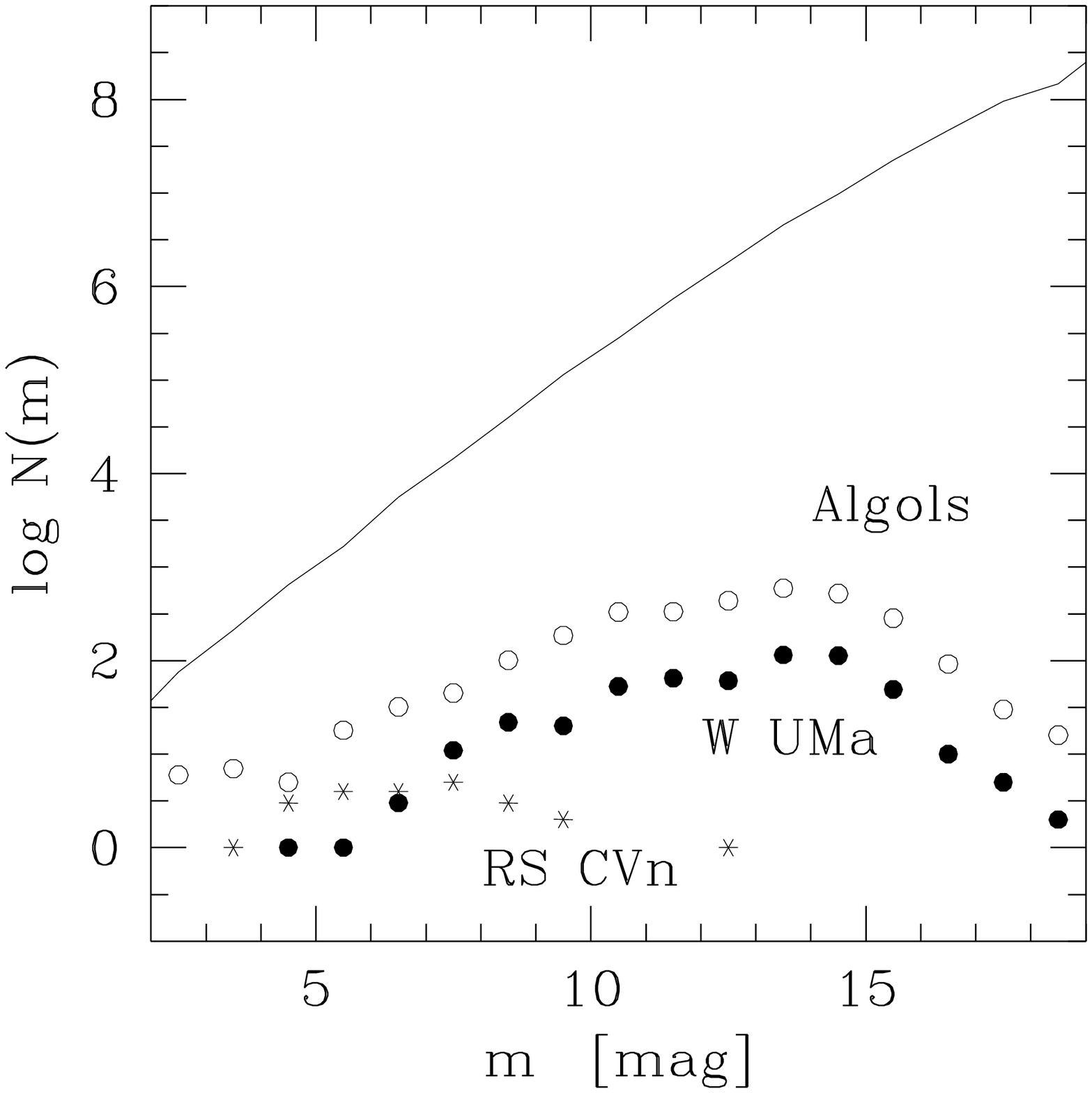}
\caption{\small
The number of all stars in the whole sky in one magnitude bins is shown
with a solid line.  The number of Algol type binaries listed in the General
Catalog of Variable Stars is shown with open circles.  The number of contact 
binaries (W UMa stars) and of RS CVn (spotted) binaries is shown with filled 
circles and with star symbols, respectively.  Note that for $ m > 9 ~ mag $
the fraction of all stars which are either Algol or contact type binaries
decreases rapidly, presumably because of incompleteness of the variable star
catalog.  The incompleteness of RS CVn type binaries becomes obvious already
for $ m > 6 ~ mag $.
}
\vspace{8cm}
\includegraphics{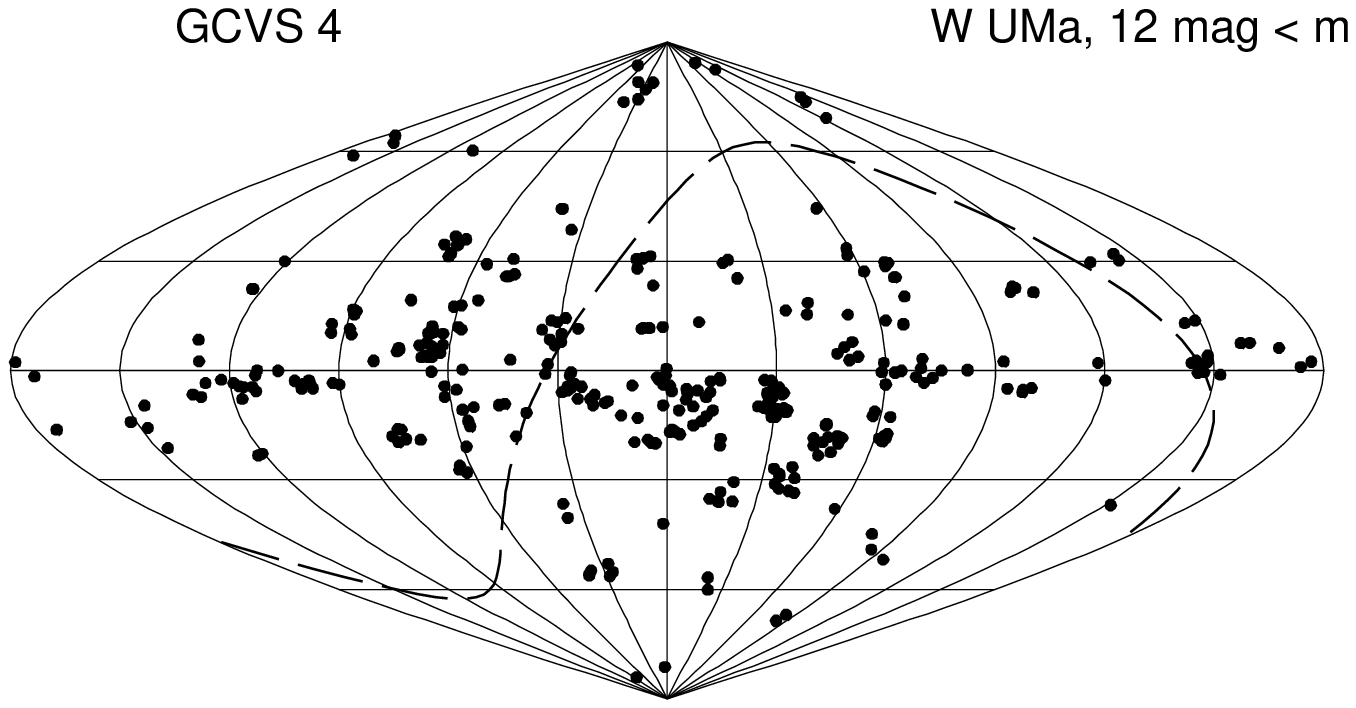}
\caption{\small
The distribution of known contact binaries (W UMa stars) which are fainter
than $ 12 ~ mag $ is shown in galactic coordinates.  
The celestial equator is marked with a dashed line.
The patchy distribution of known systems
is a clear indication of catalog's incompleteness.
}
\end{figure}

The nearly sinusoidal light variations are believed to be caused by
rotation, with the stellar surface covered with large, irregular
spots, similar to those found on the sun, but much larger.  The
spot activity varies with time, possibly in a way similar to the solar
cycle.  These red subgiant stars have hot winds, with strong X-ray emission.
It is thought that this activity is due to relatively rapid rotation
combined with convective
envelops.  There are related stars of the FK Comae type which are single,
but rapidly rotating.  These were presumably produced by mergers of 
two components of close binaries.  Photometrically it is difficult
to distinguish these two types of variables, unless the presence of
eclipses reveals the binary nature, and therefore the RS CVn classification.

Long term monitoring of
RS CVn and FK Com type variables would provide information about stellar
magnetic cycles, and so it would help us understand the nature of the
solar cycle.

\subsection{Contact Binaries}

Contact binaries, known as W Ursae Majoris stars are very common.
Recent studies
indicate that one out of 140 solar type stars is a member of a contact 
binary (Ruci\'nski 1994).  The Fourth edition of the General Catalog of
Variable Stars (Kholopov et al. 1988) lists a total of over 28,000 
variables, only 561 of them classified as EW type (i.e. W UMa type); 530 
of these have binary periods in the range 0.2 - 1.0 days.  The number
of these stars is shown as a function of their magnitude in Figure 1.
Also shown is the estimate of the total number of all stars in the
sky according to Allen (1973).   At the bright end, for $ m < 9 ~ mag $,
there is one known contact binary per $ ~ 10^3 $ stars.  Below 
magnitude 9 the fraction of known contact binaries declines rapidly.
Therefore even at $ \sim 10 ~ mag $ we may expect many new W UMa
systems to be discovered.  

The incompleteness of current catalogs
becomes striking at $ \sim 12 ~ mag $, as clearly shown in Figure 2,
presenting the distribution of those bright contact binaries in the sky in 
the galactic coordinate system.  The clustering is caused by the 
non-uniform sky coverage by the past searches.

Contact binaries are very easy to find.  Their brightness
changes continuously with a typical amplitude of $ \sim 0.6 ~ mag $.
A few dozen photometric measurements are enough to establish the
period and the type of variability.  The fraction of stars which
are contact binaries is likely to be reasonably constant well below
the 9th magnitude.  A simple inference from Figure 1 is that there
are likely to be $ \sim 10^3 $ new contact binaries to be discovered
which are brighter than 14 mag.  If the search is done with a well
defined procedure which can be calibrated then these $ \sim 10^3 $ systems
would be very valuable for a variety of studies.

The origin, structure and evolution of contact binaries is poorly
understood.  It will be very important to find out if there
are non-contact binaries which might be precursors of contact
systems, as expected theoretically.
In some theories the thermal evolution of contact systems
should periodically take them out of contact, dramatically
changing the shape of their light curves, but it
is not known if these theories are correct.  Only observational
evidence can solve this and many other puzzles of W UMa binaries.

It is very likely that contact systems, once properly understood,
will turn out to be good ''standard candles'', and as such they could
be used as tracers for studies of the galactic structure.
There is already evidence for a fairly good period - color - luminosity
relation (Ruci\'nski 1994).

\subsection{Algols}

Variable stars of the Algol type are eclipsing binaries.  They are
most often of the semi-detached type, i.e. one component fills its
Roche lobe and the gas flows from its surface towards the second
component under the influence of tidal forces.  The incompleteness
of the current catalogs can be appreciated in many ways.  First,
the fraction of stars which are known to be Algols declines as
a function of apparent magnitude, as clearly seen in Figure 1.
Second, the apparent distribution of moderately bright Algols, in
the magnitude range 12-13, is clearly clustered in the sky.
Finally, the distribution of Algols in the eclipse depth - binary
period diagram  shows a dramatic difference
between the bright ($ m < 5 ~ mag $), and moderately bright
stars ($ 9.5 < m < 10.0 ~ mag$), with the long period systems
missing from the fainter group - this is most likely caused by
the difficulty of detecting eclipses which are spaced more than a few months
apart.  Also, eclipses shallower than $ 0.25 ~ mag $ dominate the bright
sample but are entirely missing from the fainter sample,
because photographic searches could not reliably detect
low amplitude variations.

\subsection{Detached Eclipsing Binaries}

In contrast to contact binaries and most Algols the detached eclipsing
binaries are much more difficult to find as their eclipses are very narrow,
and their brightness remains constant between the eclipses.  In detached
systems both stars are much smaller than their separation.  Typically
$ \sim 300 $ photometric measurements have to be made to establish
the binary period, and roughly one out of $ 10^3 $ stars
is a detached eclipsing system (Ka\l u\.zny et al. 1995b, 1996a,b).
The fraction of such binaries which are missing in the
catalogs is likely to be even larger than it is for contact systems.

Detached eclipsing binaries are the primary source of information
of stellar masses, radii and luminosities (Anderson 1991, and references
therein).  When properly calibrated they are to become the primary
distance and age indicators (Guinan 1996, Paczy\'nski 1996a,d, and references
therein).  The first such systems were recently discovered in the Large
Magellanic Cloud (Grison et al. 1995) and in a few globular and 
old open clusters (Ka\l u\.zny et al. 1995b, 1996a,b).  This will make it
possible to measure directly stellar masses at the
main sequence turn off points in those clusters, thereby leading to
more reliable age estimates than those currently available.
The on-going searches for detached eclipsing binaries in galaxies
of the Local Group will lead to accurate determination to their
distances.  However, all these very important tasks will require
very good calibration which is possible to do only for the nearby,
and therefore apparently bright systems.  A discovery of any new bright
detached eclipsing binary makes the calibration easier and more reliable.
The brightest system of this class is $ \beta $ Aurigae (Stebbins 1910).
It has a period of 3.96 days, and the amplitude of less than $ 0.2 ~ mag $,
and at $ m \approx 2 ~ mag $ it is one of the brightest stars in the sky.

\subsection{Pulsating Stars}

Pulsating stars vary continuously, so their periods are easy to establish.
There are many different types, the best known are long period variables
(Miras), Cepheids (population II cepheids are also known as W Virginis
stars), and RR Lyrae.  Their periods are typically months, weeks, and
$ \sim 10 $ hours, and their amplitudes are a few magnitudes, somewhat
in excess of 1 magnitude, and somewhat less than 1 magnitude, respectively.

The number of these variables declines rapidly for $ m > 15 ~ mag $,
probably due to reaching the limit of our galaxy.  However, the 
incompleteness seems to set in already around $ m \sim 10 ~ mag $.
The clumpiness in the sky distribution is another indicator that
many objects are missing.   This shows strikingly in Figure 3.
The two square regions with the 
majority of RR Lyrae stars catalogued in this general direction
are 5 degrees on a side - this is a size of an image taken
with a Schmidt camera.  The apparent distribution of the RR Lyrae 
variables reveals the type of instrument used in the searches.

All types of pulsating variables are very good standard candles,
useful for distance determination, for studies of the galactic structure,
and studies of stellar evolution.  All would benefit from
complete inventories of those variables.


\begin{figure}[t]
\vspace{11cm}
\includegraphics{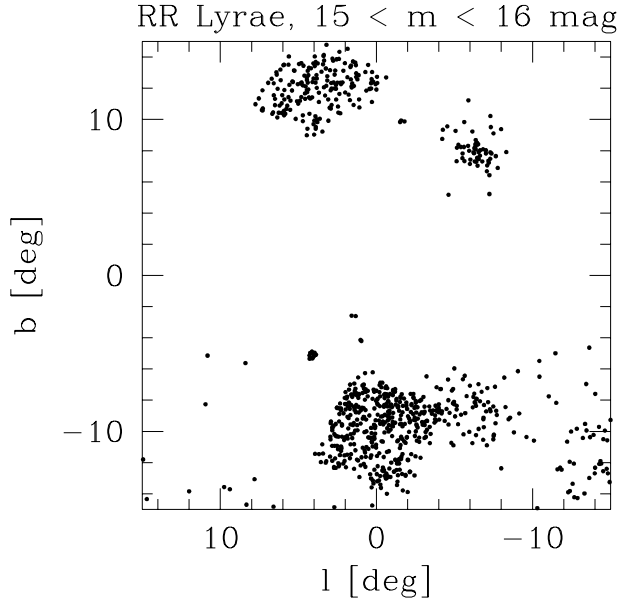}
\caption{\small
The distribution of known RR Lyrae type pulsating stars
in the magnitude range
$ 15 < m < 16 ~ mag $ is shown in galactic coordinates.  
The patchy distribution of known variables
is a clear indication of catalog's incompleteness.
}
\end{figure}

\subsection{Novae and Dwarf Novae}

Among the many types of cataclysmic variables the novae and dwarf novae
are best known.  These are binary stars with orbital periods shorter
than one day, with one star being a main sequence dwarf transferring
mass to its white dwarf companion.  Novae explode once every $ 10^3 - 10^4 $
years (theoretical estimate) as a result of ignition of hydrogen
accumulated on the white dwarf surface.  The amplitudes of light
variation are in the range 10-20 magnitudes, and the stars remain
bright between a week and a year.  The source of energy is nuclear.

Dwarf novae brighten
once every few weeks to few years, with the amplitude ranging from 3 to
7 magnitudes, and the stars remain bright between a few days and two
weeks.  The increases in brightness are cause by the enhanced viscosity
in the accretion disk around the white dwarf component, and the source
of energy is gravitational.  

The sky distribution of novae and dwarf novae
is clumpy, indicating incompleteness.  But there is more than that.  Every
year new stars are discovered to explode, so the search for new events
has no end.  The more explosions we observe, the fewer we miss, the
better our understanding of the nature of these stars, their origin and 
their evolution.

\subsection{Supernovae Type Ia}

A supernova explosion is the end of nuclear evolution of a massive
star.  This is a very spectacular but very rare event, with the typical
rate of about one explosion per century in a galaxy like ours.
Among many types of supernovae those 
of type Ia are most useful as cosmological probes.
They are standard candles with peak magnitudes
$$
m_{B,max} \approx 19.1 + 5 \log (z/0.1)  = 17.1 +5 \log (z/0.04) ,
$$
(Branch and Tammann 1992, and references therein).
Most of the scatter may be
removed using the correlation between the absolute peak magnitude and 
the initial rate of decline (Phillips 1993).  A further result has been
obtained by Riess et al. (1995a, 1996), and Hamuy at al. (1995),
who found that SN Ia light curves form a well ordered one parameter
family, with somewhat different peak luminosities (range about 0.6 mag)
and shapes.  Even the exponential declines have somewhat different slopes.
This work seems to indicate that the scatter in the Hubble diagram
of SN Ia in the redshift range $ 0.05 \leq z \leq 0.10 $ can be reduced
down to 0.1 mag in V band.

The rate of SN Ia is approximately 0.6 per $ 10^{10} ~ h^{-2} ~ L_{B, \odot } $
per century (Table 8 of van den Bergh and Tammann, 1991).  The luminosity
density in the universe can be estimated with the CfA redshift surveys
(de Lapparent, Geller and Huchra 1989) to be
$ 0.8 \times 10^8 ~ L_{B, \odot } ~ h ~ Mpc^{-3} $.  These two numbers can be
combined with Eq.~(1) to obtain the SN Ia rate for the whole sky:
$$
N_{_{SN ~ Ia}} \approx 300 \times 10^{0.6(m_{max} -17)} ~ yr^{-1} .
$$
As a large fraction of the sky cannot be monitored being too close
to the sun, and the weather is never perfect, the maximum effective 
detection rate is likely to be a factor $ \sim 2 $ lower than that
given with the eq. (2).  Still, if an all sky variability survey can
reach magnitude 17 then over a hundred type Ia supernovae will be discovered
every year, providing excellent data to improve the Phillips, 
Riess et al. and Hamuy et al. relation, and will allow even more
accurate study of the large scale flows (Riess et al. 1995b).
Also, such a survey would
provide a steady stream of alerts of supernovae prior to their
maximum brightness, allowing the most detailed follow-up studies
with the HST, the Keck, and other large telescopes.

\subsection{Quasars and other Active Galactic Nuclei}

A large fraction of active galactic nuclei (AGN) is variable, some with
very large amplitudes.  One of the most efficient ways to discover
new quasars is a search for variable objects (Hawkins and V\'ernon 1993).
There are 12 active galactic nuclei listed as variable stars in
the General Catalogue of Variable Stars (Kholopov et al. 1988).
These were first found and catalogued as variables stars, and
subsequently found to be at cosmological distances.  The following
is the list, with the observed range of magnitudes given in brackets:
AU CVn (14.2 - 20.0), W Com (11.5 - 17.5), X Com (15.9 - 17.9), GQ Com 
(14.7 - 16.1), V1102 Cyg (15.5 - 17), V395 Her (16.1 - 17.7), V396 Her 
(15.7 - 16.7), BL Lac (12.4 - 17.2), AU Leo (17. - ), AP Lib (14.0 - 16.7),
UX Psc (16. - ), BW Tau (13.7 - 16.4).  No doubt there are many other
bright and variable active galactic nuclei in the sky, and new objects
appear all the time.  A search for new AGNs as well as a continuous
monitoring of those which are already known  is very important for our 
understanding of these enigmatic objects.

\subsection{Gamma-ray Bursts}

One of the main driving forces in the plans for massive photometric
searches has been the desire to find optical counterparts (optical
flashes) associated with gamma-ray bursts (GRBs, cf. a chapter: 
{\it Counterparts - General}, pages 382-452 in Fishman et al., 1994).
This is a very ambitious undertaking.  There are two broad approaches.
One may wait for a trigger signal from a GRB detector, like BATSE, which
provides the exact time and an approximate direction to look at.  An
alternative is to have a wide field non-stop monitoring of the sky
and look later into possible coincidences with GRB detections.  In either
case the instrument and the data it will generate can be used for the
searches of all kinds of astronomical objects as described in this paper.

If a system is to work in the first mode then it would be best to have
a GRB detection system which could provide reasonably accurate positions (say
better than a degree) in real time for {\it all} strong GRBs, as these are
the most likely to have detectable counterparts.  Unfortunatelly, BATSE
detects only $ \sim 40\% $ of strong bursts and provides instant positions
good to $ \sim 5-10^o $ (cf. Fishman et al. 1992, and references therein).
It would be ideal to have small GRB detectors, like the one on the 
Ulysses planetary probe (Hurley et al. 1994), placed on a number of
geo-stationary satellites.  Such instruments could provide real time
transmission of the information about every registered $ \gamma $ photon,
and with a time baseline between the satellites of $ \sim 0.2 $ seconds 
the positions good to a $ \sim 1^o $ could be available in real time for
almost all strong bursts.  Such a good GRB alert system would put very
modest demands for the optical follow up.

In the other extreme, a blind search for optical flashes from GRBs, the
demand for the optical system capabilities are very severe, well in 
excess of any other project mentioned in this paper.  It seems reasonable 
to expect that the very powerful system that may be required, like the TOMBO 
Project (Transient Observatory for Microlensing and Bursting Objects, Otani
et al. 1996), would start small and gradually expand to the data rate
$ \sim $ terabyte per hour, along the way addressing most topics presented
in this paper.

\subsection{Killer Asteroids}

While it is possible that global disasters, like the extinction of
dinosaurs, may be caused by impacts of large asteroids,
such events are extremely rare (Chapman and Morrison 1994, and
references therein).
On the other hand smaller asteroid or cometary impacts which
happen every century may be of a considerable local concern.  The
best know example is the Tunguska event (Chyba et al. 1993, and
references therein).  In such cases there is no need to destroy
the incoming ``killer asteroid'' in outer space, it is sufficient
to provide an early warning of the impact and evacuate the site.

The less devastating events are much more common.  There are
several multi-kiloton explosions in the upper atmosphere when
small asteroids or large meteorites disintegrate, producing very
spectacular displays which are harmless (cf. Chyba 1993, and references
therein).  With a sufficiently early warning such events could be
observed and they could even provide considerable entertainment,
like the impact of the comet Shoemaker-Levy
on Jupiter in the summer of 1994, and the bright comet
Hyakutake in the spring of 1996.  

An early warning system detecting not so deadly `killer asteroids',
or rather cosmic boulders, may be feasible and inexpensive.
A mini-asteroid with
a diameter of 35 meters ($10^{-5} $ of our moon diameter) would
appear as an object of $ \sim 13 $ magnitude while at the distance
of the moon, in the direction opposite to the sun.
Moving with a typical velocity of $ \sim 10 ~ km ~ s^{-1} $
it would reach earth in $ \sim 10 $ hours.  Close
fly-byes would be far more common, and 
such events are currently detected with the Spacewatch program
(cf. Rabinowitz et al. 1993a,b, and references therein).  
If the relative transverse velocity of the cosmic boulder with
respect to earth is
$ 10 ~ km ~ s^{-1} $ then at the distance of the Moon it corresponds
to the proper motion of $ \sim 5'' ~ s^{-1} $.  Of course, it the object
is heading for earth then the proper motion is much reduced, as
the motion is mostly towards the observer.

According to Rabinowitz (1993a, Fig. 12) one boulder with a diameter
of 30 meters collides with earth once per year, and the more common
10 meter boulders do it ten times a year.  The cross section to
come to earth as close as the moon, i.e. within 60 earth radii
is larger by a factor $ \sim 60^2 = 3,600 $.  Therefore, on any given day
we may expect 10 boulders of 30 meter diameter and 100 boulders of
10 meter diameter to pass closer to us than our Moon.  At their closest 
approach these are brighter than 13 and 15.5 magnitude, respectively.
There may be dozens of nearby cosmic boulders brighter than 16 magnitude
at any time.  They are the brightest when looked at in the anti-solar
direction.  If a fair fraction of these could be detected and
recognized in real time they would offer a fair amount of excitement.
And we would learn about inhabitants of the solar system as well.

Recently, a $ \sim 300 $ meter diameter asteroid
was detected at the distance $ \sim 450,000 $ kilometers
(Spahr 1996, Spahr \& Hegenrother 1996).
It was expected to be the closest to us on May 19.690, 1996 UT,
and to be 11th magnitude at that time.

\subsection{Other Planetary Systems, Dark Matter}

The first extrasolar planetary system with a few earth-mass planets
has already been discovered (Wolszczan and Frail 1992, Wolszczan 1994).
However, this is considered peculiar, with the planets orbiting a
neutron star.  A number of super-Jupiter planets were also found around
a few nearby solar-type stars: 51 Peg (Mayor and Queloz 1995), 70 Vir
(Marcy and Butler 1996), and 47 UMa (Butler and Marcy 1996).
No doubt a detection of earth-mass planets around solar-type stars
would be very important.  The only known way to conduct a
search for earth-mass planets
with the technology which is currently available is through
gravitational microlensing (Mao and Paczy\'nski 1991, Gould and Loeb 1992,
Bennett and Rhie 1996, Paczy\'nski 1996a, and references therein).

This project requires a fairly powerful hardware, $ \sim 1 $ meter class
telescopes, and it has to be targeted in the direction where microlensing
is known to be a relatively frequent phenomenon, i.e. the galactic bulge.
It is not know how large area in the sky is suitable for the search,
and how many stars are there detectable from the ground.  An reasonable
estimates are $ \sim 100 $ square degrees and up to $ \sim 10^9 $ stars.
There are various approaches proposed.  My preference would be to
look for high magnification events with the amplitude of up to
1 magnitude and a duration of $ \sim 1 $ hour.  This would call
for a continuous monitoring program in order to acquire a large number of
photometric measurements well covering short events.  If we assume that 
every star has one earth-mass planet then the so called optical
depth to microlensing by such planets would be $ \sim 10^{-11} $,
and it would take $ \sim 100 $ hours of continuous photometric monitoring of
$ 10^9 $ stars to detects a single planetary microlensing event, i.e.
up to 20 such events could be detected every year from a good ground
based site.  Clearly, this project is very demanding in terms of data 
acquisition and data processing.

Such a search could
also either detect dark matter with compact objects in the mass range
$ \sim 10^{-8} - 10^6 ~ M_{\odot} $ (Paczy\'nski 1996 and references
therein), or place very stringent upper limits.  


\subsection{Local Luminosity and Mass Functions, Brown Dwarfs}

In general, gravitational lensing provides only statistical information
about the masses of lensing objects (Paczy\'nski 1996b).  However,
any very high proper motion star must be nearby, and hence its distance
can be measured with a trigonometric parallax.  For given stellar
trajectory it is possible to predict when the star will come close 
enough to a distant source (that is close in angle, in the projection 
onto the sky)
to act as a gravitational lens.  If the microlensing event can be
detected either photometrically (Paczy\'nski 1995) or astrometrically
(Paczy\'nski 1996c) then the mass of the lens can be directly measured.
The only problem is that in order to have a reasonable chance for
a microlensing event the high proper motion star must be located in
a region of a very high density of background sources, i.e. within the
Milky Way.  A search for such objects is very difficult because of
crowding.  However, once the rare high proper motion objects are
found the measurement of their masses by means of microlensing is a
fairly straightforward process, as such events can be predicted ahead
of time, just like occultations of stars by the known asteroids.

Recent discovery of very faint nearby objects, most likely field 
brown dwarfs, indicates that there may be a significant population
of sub-stellar objects in the galactic disk (Hawkins et al. 1996).
A discovery of such objects in the Milky Way would offer a possibility
to measure their masses by means of gravitational microlensing.
This project, just as the one described in the previous sub-section,
is very demanding in terms of data acquisition and data processing rate.

\section{Implementation}

The searches of various objects described in the previous section
cover a very broad range in the required instrumentation.  Some,
like the search for variables stars brighter than $ 13 ~ mag $, can
be conducted with a telephoto with a CCD camera attached to it.  Of
course, in order to cover the whole sky with such a search dozens
or even hundreds of such simple instruments may be needed.

The searches for nearby asteroids do not demand much larger apertures,
as many of these are expected to be brighter than $ 13 ~ mag $.  However,
as they move rapidly a very efficient data processing would be required
in order to notice them before they are gone.

The searches for variable stars are useful at any magnitude limit,
as the current catalogs are not complete, except (perhaps)
for the brightest $ \sim 1000 $ stars.  But some variables are likely
to be faint, like supernovae, as they are very far away.  It is unlikely
that a useful supernova search can be conducted with an instrument
with a diameter smaller than about 20 cm.  Any project involving a search
for microlensing events calls for a $ \sim 1 $ meter telescope.

A major technical issue facing any massive search is data processing.
The experience of the current microlensing searches demonstrated that
robust software can be developed to handle billions of photometric
measurements automatically (e.g. Pratt et al. 1996).  So far such
software runs on workstations.  However, the today's personal computers
are as powerful as yesterday's workstations, or as supercomputers
used to be.  Therefore, there is no problem in principle to transfer
the know-how to the level of serious amateur astronomers.

Once the local data processing is under control the second major problem
is the communication: how to make those gigabytes (or soon terabytes)
available to the world?  Clearly, the Internet is of some help, but not
at this volume, or at least not yet.  The problem of effective distribution 
of the vast amount of information collected in modern microlensing searches
has not been solved yet.  No doubt the solution will be found some day,
hopefuly before too long.

Some steps have already been taken on the road
towards this brave new world of massive all sky searches and monitoring.
For example the
EROS, MACHO and OGLE collaborations provide up-to-date information 
about their microlensing searches and other findings,
and a complete bibliography of their work on the 
World Wide Web and by anonymous ftp. 

There are other projects under way, some of them active for
a long time, which are now accessible over Internet.
The members of the American Association of Variable Star Observers (AAVSO)
were monitoring a large number of variable stars for many decades.
The organization is publishing an electronic journal AAVSO NEWS FLASH.
An important electronic news system is the Variable Star NETwork (VSNET).
Another Internet-based organization: The Amateur Sky Survey
(TASS), has the explicit aim to monitor the whole sky with CCD 
detectors, and to provide full access to all data over the Internet.

Yet another fascinating on-line demonstration what a modern technology can do
when combined with a human ingenuity is provided by ``Stardial'', set
by Dr. Peter R. McCullough at the roof
of the astronomy building on the campus of the University of Illinois
at Urbana-Champaign.  Stardial is
a stationary weather-proof electronic camera for
recording images of the sky at night autonomously.  It is
intended for education, primarily, but it may be of interest to
astronomers, amateur or professional, also.
At the Stardial you will find the growing archive of the data
coming from the 8x5 degree field of view camera. The limiting stellar
magnitude is $ \sim 12.5 $, through an approximately R filter bandpass.

The links providing access to all the systems mentioned above can
be found at: \\
\indent \indent \indent \indent \indent
http://www.astro.princeton.edu/\~\/richmond/surveys.html \\
\noindent
No doubt there are many more groups which are already active, or which
are planning massive photometric and/or astrometric searches, and which
communicate over the Internet.  If you know of any
other sites, or groups please let us know, and send e-mail to: \\
\indent \indent \indent \indent \indent
bp@astro.princeton.edu  \hskip 1.0cm (Bohdan Paczy\'nski), \\
or to: \\
\indent \indent \indent \indent 
richmond@astro.princeton.edu  \hskip 1.0cm (Michael Richmond).

\vskip 0.3cm

While the number of new searches increases rapidly, and so does the
volume, diversity and quality of data, there are
many challenges and many unsolved problems in the areas
of data acquisition, processing, archiving, and distribution.
The volume of data at some sites is already many terabytes,
so there is a need to develop efficient and user friendly ``search engines''.
There is a demand for new scientific questions which can be asked 
in the world of plentiful data.  The learning curve is likely to be
very long.  The full power of this new approach to observational
astrophysics will be unleashed if monitoring of the whole sky to
ever fainter limits, and ever more frequently, can be sustained for an
indefinite length of time.  However, for that to be possible
very inexpensive answers have to be found to all aspects of these
projects.  Over the years many wonderful programs had
been discontinued for the lack of funds.  I am optimistic.  Just
a few examples given above show the magic of the Internet, and they
also demonstrate that ingenious people are more important
than big budgets.

\section{Acknowledgments}

I am very grateful to Dr. Michael Richmond for setting up the WWW page
with the links to the information about many on-going massive variability
searches.
This work was supported by the NSF grants AST-9313620 and AST-9530478.

{}

\end{document}